\documentclass[10pt,conference]{IEEEtran}
\usepackage{mathptmx} % This is Times font

\usepackage{fancyhdr}
\usepackage[normalem]{ulem}
\usepackage[hyphens]{url}
\usepackage[sort,nocompress]{cite}
\usepackage[final]{microtype}
\usepackage[keeplastbox]{flushend}

\usepackage{amsmath,amsfonts,amsthm,amssymb}
\usepackage{algorithmic}
\usepackage{graphicx}
\usepackage{textcomp}
\usepackage{xcolor}
\usepackage{subcaption}
\usepackage{color,soul}
\usepackage{listings}             % Include the listings-package
\usepackage[compact]{titlesec}
\usepackage{paralist}
\usepackage{multirow, makecell}
\usepackage{nccmath}
\usepackage{fancyhdr}
\usepackage{cite}
\PassOptionsToPackage{bookmarks={false}}{hyperref}

\usepackage[hyphens]{url}
\usepackage[bookmarks=true,breaklinks=true,letterpaper=true,colorlinks,citecolor=blue,linkcolor=blue,urlcolor=blue]{hyperref}

\titlespacing{\section}{0pt}{1ex}{0.5ex}
\titlespacing{\subsection}{0pt}{1ex}{0.5ex}
\titlespacing{\subsubsection}{0pt}{0.5ex}{0.5ex}

\def\BibTeX{{\rm B\kern-.05em{\sc i\kern-.025em b}\kern-.08em
    T\kern-.1667em\lower.7ex\hbox{E}\kern-.125emX}}

\title{Machine Learning for Microprocessor Performance Bug Localization} 

\author{
Erick~Carvajal~Barboza\IEEEauthorrefmark{1},
Mahesh~Ketkar\IEEEauthorrefmark{2},
Michael~Kishinevsky\IEEEauthorrefmark{2},
Paul~Gratz\IEEEauthorrefmark{3}, and
Jiang~Hu\IEEEauthorrefmark{3} \\

\IEEEauthorrefmark{1}Universidad de Costa Rica, 
\IEEEauthorrefmark{2}Texas A\&M University, 
\IEEEauthorrefmark{3}Intel Corportation \\

Corresponding Author: erick.carvajalbarboza@ucr.ac.cr
}

\renewcommand{\baselinestretch}{0.95}

\begin{document}
\maketitle
\thispagestyle{plain}
\pagestyle{plain}

%%%%%% -- PAPER CONTENT STARTS-- %%%%%%%%
\begin{abstract}
The validation process for microprocessors is a very complex task that
consumes substantial engineering time during the design process.
Bugs that degrade overall system performance, without affecting its functional 
correctness, are particularly difficult to debug given the lack of a
golden reference for bug-free performance. 
This work introduces two automated performance
bug localization methodologies based on machine learning
that aim to aid the debugging process.  Our results show that, for the evaluated
microprocessor core performance bugs whose average IPC impact is
greater than 1\%, our best performing technique is able to localize
the exact microarchitectural unit of the bug $\sim$77\% of the time,
while achieving a top-3 unit accuracy (out of 11 possible locations)
of over 90\% for bugs with the same average IPC impact. 
The proposed system in our simulation setup requires only a few seconds
to perform a bug location inference, which leads to a  
reduced debugging time.

\end{abstract}

\section{Introduction} \label{sec:intro}

Large amounts of time and effort are devoted to
verification and validation of every microprocessor design project.
Broadly, design verification can be broken into two large categories:
(1) functional and (2) performance verification, which is to identify design bugs that degrade performance without affecting functionality. Performance bugs are different from performance bottleneck as the former is due to design mistakes while the later is caused by tight resource constraints. Performance loss due to performance bugs  can 
be very significant, with recent reported cases shown to be
$>10\%$~\cite{mccalpin2018hpl}. This demonstrates a critical 
need for automated mechanisms for performance debugging.  As 
recent designs from Intel~\cite{corei7-11}, AMD~\cite{ryzen-9},
ARM~\cite{cortex-a}, and others place an even greater emphasis
on core performance, design complexity has scaled
dramatically, likewise scaling the difficulty in all forms of
verification.

%Functional verification has received extensive attention from researchers and, although complex, it benefits from the availability of known correct outputs that can be used to compare against.

Performance verification at microarchitecture level ensures that a
design correctly achieves expected performance in terms of execution
time or cycle count.  The main challenge in this task is that, unlike
functional verification, there is no exact golden reference to compare
against.  This is because of the high difficulty of modeling all the
interactions between the different units in complex microprocessor
designs, and accurately represent how they affect the overall system
performance.  %This task also suffers from 
%the lack of a good debugging infrastructure, as well as from 
%limited visibility into intermediate points in the design, which are mostly exposed through performance counters. Although useful for estimating the performance of the system, these counters are very difficult to use for manual debugging because of their complex relationship with processor performance and due to the large amounts of data they generate.  
Traditionally, performance
verification is conducted mostly through manual techniques which rely
on rough estimations of performance gain expected by
microarchitectural changes~\cite{Singhal2004}. Such manual processes
are not only very lengthy but also error-prone.

The process of performance verification and debugging roughly consists of two steps: (1)~detection, which determines whether a
design achieves expected performance or not, and (2)~localization,
which identifies the microarchitectural units causing the performance
issues and is the focus of this work.

There are few previous studies on automating detection of
microprocessor performance bugs~\cite{Bose1994,
  surya1994architectural,carvajal2021detection}. 
The majority of those~\cite{Bose1994, surya1994architectural} relies on capturing
design intentions using a bespoke performance model as a golden
reference, this  entails long development time and may contain
errors by itself. Recently, a data driven and machine learning
(ML)-based approach~\cite{carvajal2021detection} was developed for
automatic performance bug detection with high accuracy. Although
significant, these works do not solve the 
problem of performance bug localization.
%pressing problem of identifying where the performance bug is.

Works in automating microprocessor performance bug localization
are even scarcer.  Adir \emph{et al.}~\cite{adir2005generic}
propose perhaps the only partially related work of which we are
aware.  Their work focuses on formal planning of test program
generation for individual units, such as issue queues. This strategy
follows conventional functional verification, involving heavy
manual effort, costing significant engineer-time to develop a test
plan, and as much as ten days of computer runtime per functional
unit. To the best of our knowledge, there has been no systematic study
on automatic performance bug localization for microarchitecture
designs.

Performance bug localization is a complicated task, which is currently
solved using mostly manual techniques.
Even
in the more widely studied area of functional validation, the industry
lacks a standardized mechanism to automate bug localization, it has
been only recently that academic efforts have attempted to automate
this task~\cite{BugMD}. Considering this, it is important to note that
any type of design automation which successfully reduces the 
time and effort required by engineers to debug their designs is highly
valuable. Since automatic performance debug for microprocessors is
a huge yet under-studied challenge, it is very difficult, if not impossible, to find a perfect solution in a single work. Although our work is not perfect, it serves a key stepping stone 
 toward solving the problem.

This work tackles the performance bug localization problem by
using ML to generate a ranked list of most likely mi\-cro\-ar\-chi\-tec\-tur\-al units that 
might contain the bug.  This list may be used
to prioritize the debugging order, as well as to identify
teams with the right expertise to perform further debug. Two different methodologies are
proposed, evaluated, and contrasted. These data-driven
techniques achieve high
accuracy, while being fully automated. Further, they
consider intra- and inter-unit interactions, as opposed to other
techniques proposed in the partially related previous work~\cite{adir2005generic} which
considered only intra-unit behavior.

%Our methods are based on ML, wherein our models are
%trained using data from legacy designs.
%To take the full advantage of
%these approaches, we assume that architectural changes in a new design
%are incremental when compared to its previous
%generations. Examining recent processors from major vendors including
%Intel, ARM, and AMD, we find this assumption holds true, since the generational change in microarchitectures
%is largely incremental. Thus, the methodologies proposed here provide
%alue for a multitude of upcoming designs.  However, even when
%disruptive changes occur, the methodologies can still be beneficial for bug localization on structures that conform to previous microarchitectures, using workloads that
%do not exercise new functionalities. Further, as general purpose microarchitectures become ever more mature, and the inter-generational performance gains decrease, 
%it is even more important to retain as much performance as possible, making performance debugging ever more important.

The major contributions of this work include the following:
\begin{compactitem}
\itemsep0em 
\item This is the first systematic study on fully automatic
  performance bug localization for microarchitecture designs, to the
  best of our knowledge.

\item Two ML-based approaches to tackle performance bug
  localization, as well as a hybrid of both, are evaluated
  and contrasted.

\item For bugs with an average IPC impact greater than 1\%, our best
  performing methodology identifies the correct bug location as the
  most likely unit in $\sim77\%$ of the cases, and achieves over 90\%
  accuracy when the three most likely options (out of 11 possible) are
  considered.  

\item One of the proposed methodologies is not only very accurate localizing
performance bugs, but it can also be applied to confirm the results
of performance bug detection with high accuracy.

\item Although the focus of this work is on microprocessor core,
we evaluated our methodologies on the processor memory hierarchy. This evaluation
uses a different experimental setup, showing the robustness of the proposed techniques.

\end{compactitem}

As an early work on performance bug localization, the design of this study is subject to potential limitations, however, we feel it still represents a good first step towards solving the problem. The scope of our work and its limitations are as follows:
\begin{compactitem}

\item Legacy
designs with identified performance bugs are required, so that the ML
models can be trained. Bug-free legacy designs are required only 
in one of the methodologies, yet, if available, the other can take advantage of the additional data.
However, thanks to the thorough pre- and post-silicon
debug to which the designs are submitted, these legacy designs are
generally available.

\item We assume that only one bug is present at a time, 
in parallel to the single-fault model which is common practice
in VLSI testing works. As explained in Section~\ref{subsec:impl_bugs}, we still expect 
our methodologies to work well in the presence of multiple bugs in a single design.

\item Our methodologies do not provide a quantitative coverage guarantee.  
In general, performance bug
coverage is extremely difficult to define and is a potential
research topic on its own. We know of no prior work which presents a
definition of such coverage. Nonetheless, the evaluated bugs are based on published errata, cover a large amount of microarchitectural units and affect the system in a variety of ways. Thus, we feel these bugs represent a reasonable start for early work in this area.

\item We assume that there are no dramatic structural
microarchitectural changes between the legacy designs and the
designs under debug. Examining recent processors from major vendors, including Intel, ARM, and AMD, we find this assumption holds true, since the generational change in microarchitectures
is largely incremental. That said, even when larger shifts occur, the
methodologies can be partially reused. For example, consider the
introduction of the AVX instructions with Intel's Sandy Bridge
architecture in 2011.  Initially there would be no available data to
test these instructions using our methodologies, however the rest of
the Sandy Bridge design could be debugged with our methodology,
leveraging workloads that do not exercise the new instructions.  In
future implementations, data from Sandy Bridge can be used to build
the models required to use our methods for debugging AVX. 
%Further, as general purpose microarchitectures become even more mature, and the inter-generational performance gains decrease, it is even more important to retain as much performance as possible, making performance debugging even more important.}

\item We limit our evaluation to a pre-silicon setup, because
it is infeasible for us to inject known design bugs in silicon to
evaluate the methodologies.  Further, should our methodologies be
applied to a commercially available design, and an actual bug be
found and localized, we would not be able to verify that such
localization is correct without prior knowledge of its existence so
as to verify our findings. However, our methodologies can be applied in both pre- and post-silicon scenarios. During pre-silicon stages fixing performance 
bugs is easier and cheaper, 
the availability of performance counters is greater (due to the usage of a
simulator) and the counters can be sampled at a much faster rate. 
By using only counters available in-silicon, and adjusting the sampling frequency, we could use the proposed
methodologies during post-silicon stages. In post-silicon analysis the methodology
could be applied to longer workloads, providing a way to exercise complicated bugs that
are not possible to trigger with short pre-silicon traces.
%Further, we can follow hybrid approaches where the ML model training is performed using simulations, and the techniques are applied to data obtained from microprocessors during post-silicon debug. }

\end{compactitem}

Despite the aforementioned, we present a first, useful, yet attainable,
step towards the goal of performance bug localization, and we hope this work can draw the attention of the research community
to the broader performance validation domain.

\iffalse{
In Section~\ref{sec:scope} we describe the problem
formulation and outline the scope of this work.  We note that, to
date, very little work exists in automating performance bug
localization. 

Section~\ref{sec:methodology} 
describes the approaches developed to tackle the performance bug
localization task.  Section~\ref{sec:experimental_setup} provides
details of the architectures, and performance bugs used for
evaluation. Section~\ref{sec:evaluation} presents results obtained in
several experiments developed to evaluate the methodologies. A brief
review of previous work related to performance debugging is presented
in Section~\ref{sec:related_work}.  And finally,
Section~\ref{sec:conclusion} concludes the paper.
}
\fi

\section{Problem Definition and Scope}
\label{sec:scope}

%FORMULATION, OBJECTIVE, INPUT, ASSUMPTIONS

The \emph{objective} of this work is to identify the
microarchitectural units where a detected performance bug is located.
The \emph{scope} of this work, as presented here, is limited to
microarchitectural-level performance bugs in the processor core that
affect the IPC of the system\footnote{We note that, while we do not
test on components other than the processor core, there is no
reason to think that this methodology could not work there.}.
Bugs that might affect circuit-level timing (\emph{i.e.} clock
period) are not considered. For testing and insertion convenience, we cover cases where the hardware does not achieve the expected performance due to a microarchitecture bug, that is, due to waste or under-utilization of resources by implementation faults. The methodologies, however, can be used for performance bottlenecks arising from suboptimal hardware, algorithms, or other settings.

\section{Methodology} \label{sec:methodology}

\subsection{Overview}

In this work, we propose two machine learning-based methodologies aiming to
identify the microarchitectural unit where a detected performance
bug exists. Here, we use the term microarchitectural
unit to refer us to a segment of a microprocessor that performs a
specific task. Examples of microarchitectural units in this context
would be Fetch, Decode, Branch Predictors, Load/Store Queue, etc.
  
Both of the proposed approaches leverage performance counter data as
inputs, since they are  available in almost any
microprocessor design.  This prevents the overhead of adding dedicated
data acquisition infrastructure.  In addition, microarchitecture
designs usually have hundreds or thousands of Performance
Counters~\cite{perfmon-intel,perf_counters_amd}, which are generally
sufficient to extract necessary information for performance
estimation. 
%Where prior work has only attempted to detect the existence of such bugs~\cite{carvajal2021detection}, this work aims to determine which functional unit contains that bug.

Here, we use ML due to its capability of knowledge extraction from
data, and its strength to handle complicated nonlinear behaviors.
Existing bug localization, in practice, is by and large manual, while
ML is perhaps the best approach to mimic a human manual process among
mathematical or algorithmic options. However, our goal is not to
remove the human from the debugging task, but merely speedup the
process by providing useful guidance extracted from the data.
Performance counters are used as input features to our ML-based
methodologies.  They are extracted from the results of simulating
(or executing) a diverse set of workloads.
%, which are used as stimuli applied to the microprocessor, in order to measure how it behaves under different circumstances.

The first proposed methodology uses multiple ML models,
each of which classifies if a given unit contains the bug. Then,
results of these models are aggregated across different workloads to
obtain an ordered list of units according to their confidence levels
for bug existence. This methodology is referred to as
\textbf{``Counter-Based Classification''} or \textbf{``CBC''}.

The second methodology is a two stage approach.  Its first stage
includes machine learning models for predicting bug-free performance
in terms of IPC. In the second stage, prediction errors of these
models are utilized for estimating the likelihood of bug existence for
each unit.  The name \textbf{``Performance Prediction error-Based
  Classification''} or \textbf{``P2BC''} is used to refer to this
method.

The output from either of the methodologies provides a priority list
of the most likely microarchitecture units that might contain the
performance bug, this list can be used for further analysis by
validation or design engineers.

Similar to prior work in bug
detection~\cite{carvajal2021detection}, we leverage the use of
SimPoints~\cite{sherwood2002simpoints} in our performance bug
localization techniques, to identify orthogonal workloads from long
running applications, such as those on the SPEC CPU
suites~\cite{spec2006,spec2017}. With this, short and performance
orthogonal traces, that are relevant for microarchitecture performance
verification, can be automatically extracted. However, the
methodologies proposed here are not restricted to the usage of
SimPoints, and any workload that validation or design engineers
consider appropriate to verify the design can be incorporated and
should only improve the results, as discussed in Section~\ref{subsec:nb_probes}.

\subsection{Performance Counter Selection} \label{subsec:counter_selection}
A microprocessor
typically has hundreds or thousands of performance counters. Since using all of them makes the models unnecessarily large, a small subset is
obtained for each workload using an automated methodology that follows the two-step algorithm described below.
\begin{compactenum}
\itemsep0em 
\item The average Pearson correlation coefficient between each counter
  and the corresponding microbenchmark's IPC across multiple legacy
  architectures is calculated. Counters that are not highly correlated
  with IPC (magnitude lower than a threshold $\alpha$) are removed.
\item Correlation between each pair of the remaining counters is
  calculated. Two highly correlated
  counters (magnitude greater than $\beta$)
  will provide the model with redundant data, in that case,
  one of them is pruned from the list.
\end{compactenum}

The counter selection is completely orthogonal to the bugs that might be present on the system, the procedure is based entirely on the correlation between performance counters and 
IPC in legacy architectures. Although more sophisticated techniques  for automatic extraction of relevant performance events have been recently proposed~\cite{lv2018counterminer}, we find that Pearson correlation factor works sufficiently well in our setup.

We note that different counters are selected for each workload. Among the performance counters that are most frequently selected by our automated methodology we have the following: the number of fetched instructions, percentage of branch instructions, number of writes to registers, percentage of correctly predicted indirect branches, etc.

\subsection{Methodology \#1: Counter Based Classification (CBC)} \label{subsec:one_stage_methodology}
\begin{figure*}[tb!]
  \centering
  \includegraphics[width = 0.65\textwidth]{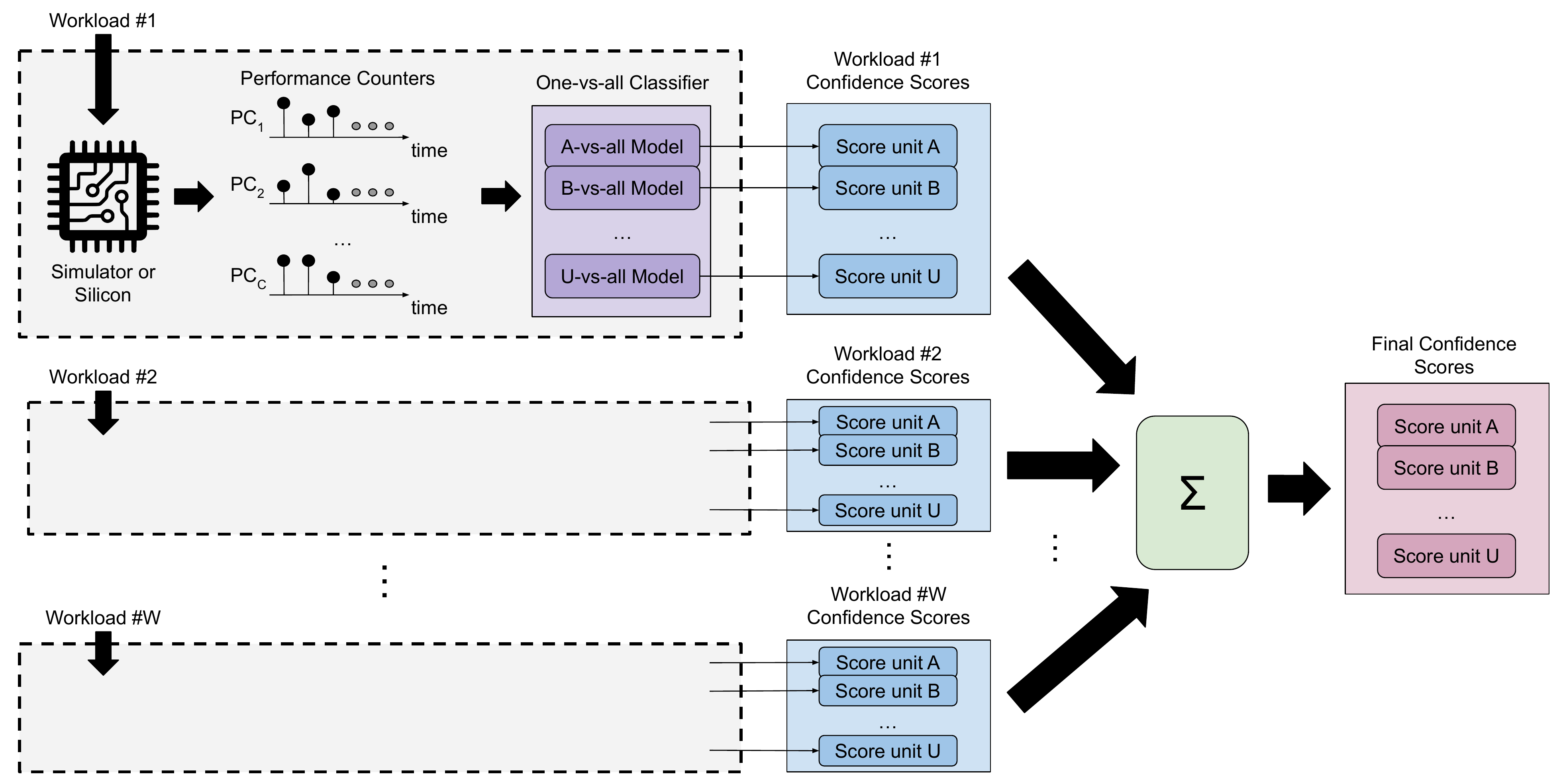}
  \caption{Overview of the CBC methodology.}
  \label{fig:one_stage_methodology}
\end{figure*}

The methodology proposed in this section consists of a single machine
learning stage followed by an aggregation procedure. An overview of
this methodology is shown in Figure~\ref{fig:one_stage_methodology}.

If the set of all the evaluated workloads is denoted by $W$ then,
for any given workload $w_i \in W$, we extract data from a set of
performance counters.  The counter data may be taken either from a
simulation or from the actual silicon chip.  The counters used are
those previously selected, using the methodology described in
Section~\ref{subsec:counter_selection}.

Although the automated counter selection can lead to a different
set of performance counters for each workload, CBC achieves better accuracy when every
model uses the super-set composed by the union of the counters
selected for each of the workloads.  This is because, for some
workloads, the selected counters do not contain any information
regarding specific units that might be affected by the performance
bug. Providing every workload with this larger counter set increases
the visibility the models have in all the units of the
system. Although using the union super-set significantly increases
the number of features per model by a factor of $\sim15\times$, this
merged set is still about $10\times$ smaller than the complete list
of available counters.

The data extracted from the performance counters is sampled and reset
every time a certain number of cycles have passed (\emph{e.g.}
every 100k cycles). Resetting the counters ensures 
that their value reflects only their
behavior in the current time-step, without keeping
track of its history. Therefore, the counter data for workload $w_i$ is
arranged as a time-series with $T_i$ elements. This time-series data is then feed into a 
ML classifier. 

In this work, we formulated the bug localization problem
as a multi-class classification task. Here, every 
microarchitectural unit where the bug
might be located corresponds to a class $u_j \in U$, where $U$ is the
set of all units.  To solve the multi-class classification task, some
common strategies in the ML community are one-vs-one (OvO)~\cite{ovo},
one-vs-all (OvA)~\cite{ova} or using a single multi-output model~\cite{bishop2006pattern}.  Although the usage of the latter
was evaluated, it did not achieve the desired
accuracy. A OvA strategy is followed, as it provides
good accuracy with a small fraction of the base classifiers that would be 
required
for a OvO strategy.  Here, a base classifier is a model to
classify if the bug exists in a specific unit for a workload.
With OvA, a total of $|W|\cdot|U|$ base models
(classifiers) are needed.

The base classifier for workload $w_i$ that flags the bug
in unit $u_j$ is denoted as $m_{i,j}$. Each $m_{i,j}$
produces a confidence score,
which is a soft classification in $[0,1]$ for workload $w_i$ to tell
if the bug is at unit $u_j$. Since the confidence scores from all
classifiers of a workload do not add up to one, they do not represent
probability. However, they serve as probability proxies, as higher
scores mean higher probability of the bug being in that unit. The
confidence scores for every unit across all the workloads are summed to
create a final score for each unit. The units with higher scores
have higher probability of the bug being present in
that unit. If sorted, the output provided by CBC represents a ranking of the
most likely units where the bug might be located.

Each base classifier $m_{i,j}$ is trained with data of workload $w_i$ from
one or more
legacy architectures. Samples from legacy architectures with bugs 
occurring at unit $u_j$ are considered ``positive'' cases for the model,
while bugs in any other unit are considered ``negative''. In this way,
the base classifier will learn to identify bugs exclusively on its 
corresponding unit. If available, samples from bug-free legacy architectures
can be used for training, and will be considered ``negative'' samples.

Because of the time-series format used for the performance counters, two different
methods to calculate the confidence score are evaluated and
contrasted:
\begin{compactenum}
\itemsep0em 
\item \emph{Per-trace classification}: Using neural networks that are
  able to take advantage of temporal locality, such as Convolutional
  Neural Networks (CNN)~\cite{lecun1995convolutional} or Long Short-Term
  Neural Networks (LSTM)~\cite{hochreiter1997lstm}, the complete time trace
  can be processed and a single score generated in the end. The
  proposed methodology uses CNNs, as LSTMs did not produce
  satisfactory results.

\item \emph{Per-time-step classification}: At every time-step $t_{i}$,
  a bug location prediction is performed by using the input features
  related to that specific time-step.  With this, a classification
  result per time-step is obtained, ultimately creating a
  ``classification trace''.  Previous time-steps could be added as
  input features in order to provide the models with information
  regarding counter history. However, we found that for the evaluated
  time-step size (100k cycles), adding these had no significant
  benefit.
  This method allows for other ML methods to be used, such as
  Multi-Layer Perceptrons~\cite{hornik1989multilayer}, Random
  Forest~\cite{breiman2001random} or Gradient Boosted Decision
  Trees (GBDT)~\cite{friedman2001greedy}.  Although all these methods were
  evaluated, the proposed methodology uses GBDT, as
  it was found the best performing.  Ultimately, to transform the
  prediction trace into a single value, the mean value
  of prediction scores
  across the
  whole time trace is used as the final score.
\end{compactenum}

\subsection{Methodology \#2: Performance Prediction error-Based Classification (P2BC)} \label{subsec:two_stage_methodology}

\begin{figure*}[htb!]
  \centering
   \includegraphics[width = 0.9\textwidth]{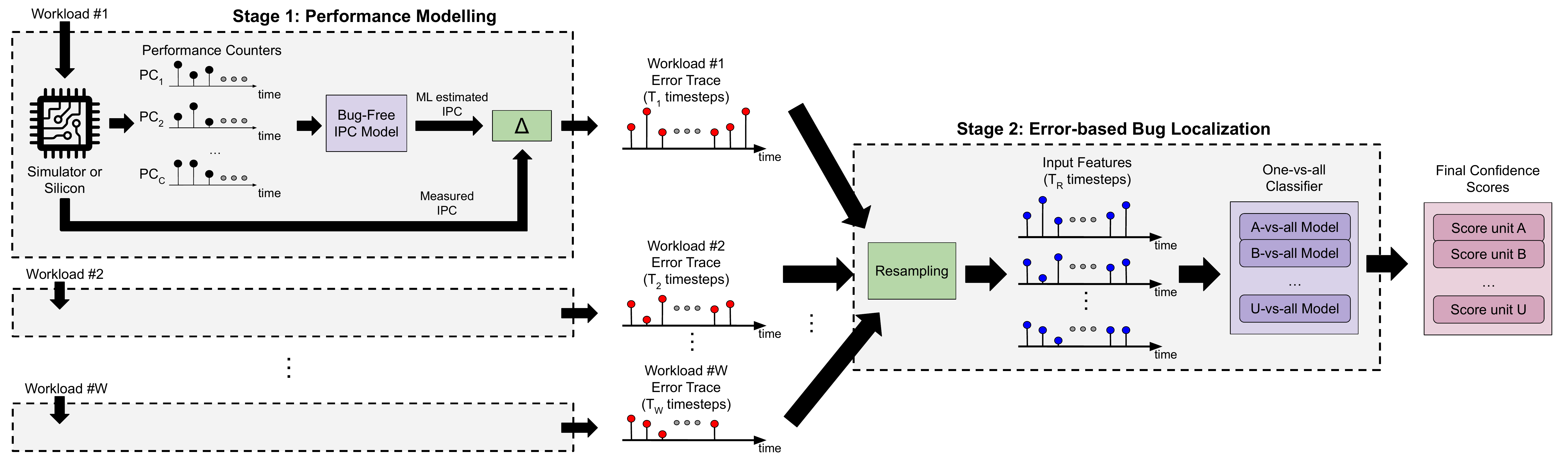}
  \caption{Overview of the P2BC methodology.}
  \label{fig:two_stage_methodology_overview}
\end{figure*}

The Performance Prediction error-Based Classification (P2BC) method is
composed of two different ML stages. An overview of this
is shown in Figure~\ref{fig:two_stage_methodology_overview}. Its first
stage is a set of performance (IPC) prediction models trained
exclusively with bug-free design data, so as to capture the
relationship between the counters and the performance under healthy
conditions.  When such models are applied on buggy designs,
significant prediction errors show up as the healthy conditions no
longer hold.  The second stage uses those prediction errors as
features for its ML classifiers. The intuition behind this approach is
that the specific workloads where the model and the ground truth diverge,
as well as the characteristics of that divergence, can provide
key information on which particular functional unit is the source of
the bug. Hence, P2BC localizes bugs according to symptoms of
performance model failures.

The first stage of the P2BC the methodology is heavily inspired by
prior work on performance bug detection~\cite{carvajal2021detection}, although the second stage of
P2BC differs significantly, as it leverages the full error trace to feed the ML models for
performance bug localization, while prior work used a single error
metric as input to a heuristic rule-based classifier merely for
determining bug existence.

\subsubsection{P2BC Stage 1: Performance Modeling}

Just as described in Section~\ref{subsec:one_stage_methodology}, we extract data from a set of performance counters, either via simulation
or from the actual silicon chip.
The goal of this stage is to model the bug-free processor
performance. To do this, we developed ML models 
that use the performance counter data as input features in order to infer
bug-free behavior of an specific target metric (IPC in this case).
There is one ML regression model for each workload, with counters selected according to
Section~\ref{subsec:counter_selection}. 

Unlike CBC where, in order to increase visibility, the super-set of
the counters selected across all the workloads are used as features
for the ML models, in P2BC each workload uses only
the counters selected specifically for it, as those are sufficient to infer the 
performance obtained by the system. Increasing the set of counters might provide 
undesired redundancy that will reduce the model sensitivity to performance bugs,
degrading the methodology's capacity to localize the bugs.

We use a per-workload model strategy since, due to their
specialization to the behavior of
each microbenchmark, it
achieves lower error (7.7\% RRMSE, in average for our setup) than workload agnostic models (27\%, in average~\cite{ardalani2015cross}). Further, our methodology does not support the usage of static analysis models~\cite{ardalani2019static}, as they are unlikely to trigger bugs due to not being executed on the actual design.

In order to train the bug-free IPC models, we use performance counter data coming exclusively
from bug-free legacy designs in order to establish a bug-free performance baseline. 
In discussions with industry partners, we
find that their availability can be safely assumed given the extensive post-silicon
debug that these designs have gone through by their end-of-life. However, even if it is not
possible to have designs that are completely bug-free, with P2BC we would be able to localize
new bugs that are not part of the baselines learned by the ML models.
That said, prior work~\cite{carvajal2021detection} demonstrated that even
when small performance bugs exist, performance prediction can perform
well.  Using different legacy designs ensures the model learns to
differentiate abnormal behavior due to performance bugs vs. due to
different microarchitectures.

Here, we use GBDT~\cite{friedman2001greedy} models
for performance prediction, which were shown to be the best 
technique for this task~\cite{carvajal2021detection}. These
performance models are not meant to be golden references, in fact,
they perform poorly for buggy designs. However, the prediction errors
contain useful information and serve as features for the 2nd stage
classifiers. Since the performance counters are processed in a time-series
manner, these models produce an inferred (or ML estimated) IPC time trace.

Once the inferred IPC trace is obtained, an error
trace is calculated by computing the difference between
the IPC obtained via ML performance models and the one obtained from
the simulations (or in-silicon execution). These error traces are used
as inputs for the second stage of this methodology.

\subsubsection{P2BC Stage 2: Error-based Bug Localization}

Since the error traces from stage one come from different workloads,
which might have different execution times,
the number of time-steps on each trace might differ between workloads. Therefore,
after error traces are obtained for all the
workloads, they go through a resampling procedure.
The goal of this procedure is to transform all the traces so that they all 
have a uniform number of time-steps $T_R$,
so that ML strategies like CNNs
can be used. This is only required due to the mixing of data from different
workloads, since that is not needed by CBC, resampling is not required in that 
method.

The resampling we use is the Fourier methodology implemented on the SciPy library~\cite{scipy}. This procedure is based on Nyquist-Shannon sampling theorem
\cite{shannon1949communication}.  The basic idea is that, by modifying
the frequency domain spectrum of a signal, the number of samples
needed to capture all the trace information can be changed as well.
Depending on the number of samples in the trace, there can be two
possible resampling mechanisms:

\begin{compactitem}
    \itemsep0em
  \item \textbf{Down-sampling}: This is when the number of 
    time-steps in $w_i$, denoted $T_i$, 
    is greater than $T_R$. In this case, after an FFT is
    applied, the frequency spectrum is truncated at the required
    maximum frequency. With this, the signal can be rebuilt with less
    samples via FFT\textsuperscript{-1}.
  \item \textbf{Up-sampling}: This is when $T_i < T_R$. Here, after an FFT is applied,
    the frequency spectrum is zero-padded.  Then, the signal can be
    rebuilt with more samples via  FFT\textsuperscript{-1}.
\end{compactitem}

Empirical results demonstrated that the best results can be achieved
when the target number of samples is equal to the average number of
time-steps across all the workloads. 

The resampled traces are used as inputs to a multi-class 
classifier which is trained to identify the
microarchitectural unit where a performance bug is located. A
``one-vs-all'' methodology is followed, same as in
Section~\ref{subsec:one_stage_methodology}.

The IPC inference errors obtained by using stage one in legacy architectures
are used to train this stage classifier. As opposed to CBC, where one classifier
per workload is needed, here we use a single OvA classifier. Each base classifier $m_{j}$
is trained with data from one or more legacy architectures, samples from legacy
architectures with bugs occurring at unit $u_j$ are considered
``positive'' cases for the model, while bugs in any other unit
are considered ``negative''. Note that the training of this stage, bug-free architectures
are not mandatory, as they are merely used as ``negative'' samples.

The classifiers are implemented using a 1D-CNN.  The convolution operations are performed exclusively
along different time-steps of the same workload, and every workload is
used as a different channel~\cite{lecun1995convolutional}. The reason
to do this is because there is no relevant information to be learned
across same time-steps of different workload given that the order in which
they appear is
completely arbitrary.

When an inference is performed, the confidence score of every model 
is proportional to the probability assigned by the model for a bug being present at each 
unit. Higher confidence scores are obtained for the most likely locations of the bug.

\subsection{Trade-offs} \label{subsec:tradeoffs}

These two methodologies have their advantages and drawbacks, these
are discussed as follows:
\begin{compactenum}
\itemsep0em 
\item \textbf{Accuracy vs data storage and runtime}: 
  As we show in Section~\ref{sec:evaluation}, CBC generally performs better than P2BC.  However, CBC
  requires a higher number of models than  P2BC. For the former, one model per unit per workload is needed
  ($|W| \cdot |U|$ in total), while the latter requires one model per
  workload for IPC estimation (Stage 1) and one model per unit for bug
  localization (Stage 2), for a total of $|W| + |U|$.  The increased number of
  models in CBC creates a larger data storage requirement
  ($100\times$ in our evaluation), as well as a longer runtime for
  training and inference ($6-10\times$).
  Nevertheless, although CBC is slower than P2BC, for both
  methodologies, full training can be achieved within
  a day, while the inference time is in the order of a just few seconds,
  without paralellization.

\item \textbf{Incremental workload addition}: CBC is more friendly to
  incrementally adding new workloads.  To add a new workload, $|U|$
  new models must be trained for this method. Although for P2BC the
  number of trained models is not significantly larger, only $|U|+1$ (one
  new IPC model, and $|U|$ re-trained models in stage two), adding a new workload
  requires to re-train the entire stage two, which could impact the
  quality of results and requires a longer training period.
    
\item \textbf{Incremental bug unit location addition}: Both approaches
  allow for incremental addition of microarchitectural units where
  bugs might be located.  This is useful for cases when a bug location
  that was not part of the initial list of considered classes needs to be included, or when a new structure is added
  to the system.  In this case, only one new model would need to be
  trained for P2BC, while CBC would need $|W|$. Although
  re-training the models for other units using designs with bugs in
  the new one is not required, doing so might improve the accuracy of
  the overall debugging approach.

\end{compactenum}

\subsection{Ensemble of both methods} \label{subsec:ensemble}

Although both methodologies provide satisfactory results, as shown in
Section~\ref{sec:evaluation}, they don't excel in the same cases.  In
order to take advantage of the strengths of both methods, a simple
ensemble scheme is presented in this section. The procedure is as
follows.

\begin{compactenum}
\itemsep0em 
\item The final confidence scores of each methodology are normalized,
  so that for every design, the sum of the confidences across all the
  units $u_j$ equals one.  Since a ``one-vs-all'' methodology is used,
  this cannot be guaranteed beforehand without this step.
\item For each unit, the average score obtained across both
  methodologies represents its final score.
\item This newly calculated scores are ranked, and their order is used
  to determine the unit with the highest probability to have a
  performance bug.
\end{compactenum}

Overall, the overhead of the proposed methodologies is very low, as ML model training takes about 30-60 minutes per model, and inference is in the order of seconds (this could be further accelerated with GPUs). Further, the simulation of each SimPoint takes 10-20 minutes to complete. Since all our models and simulations are independent, with enough computational resources, they can all be launched in parallel, producing a negligible overhead on the overall debugging process. 

\section{Experimental Setup} \label{sec:experimental_setup}

This section elaborates on the evaluation setup of the
proposed methodologies.  Since performance bug localization in cores
is the focus of this work, sections~\ref{subsec:probe_setup}
to~\ref{subsec:impl_bugs} cover in detail its experimental setup
characteristics, while section~\ref{subsec:setup_memory} covers the
changes implemented to apply the proposed methodologies to a memory
system performance bug localization setup.

\subsection{Workload Setup} \label{subsec:probe_setup}

We extracted a total of 190 SimPoints~\cite{sherwood2002simpoints}  from
10 applications in the SPEC CPU2006
benchmark~\cite{spec2006} in order to use them as the set of workloads to
evaluate the proposed methodologies. Our methods consider each of these
SimPoints to be individual workloads. Although SPEC CPU2006 is used, nothing
prevents the usage of the more recent SPEC CPU2017\cite{spec2017},
other benchmark suites, or any custom-made workload the designers or
verification engineers consider relevant for performance validation of
the design under test.  We used SPEC CPU2006 for its shorter
runtimes, smaller memory footprint and greater compatibility with the
gem5~\cite{gem5} execution environment. Unlike cases
where performance improvement techniques are evaluated, here, the
benchmarks are used to extract orthogonal workloads, so complete suite execution is not required.

Each SimPoint contains $\sim$10M instructions, and the applications
used for SimPoint extraction are an arbitrary set of 10 that
were able to compile and run successfully in gem5 across the evaluated
microarchitectures. These applications are: perlbench, bzip2, gcc,
mcf, milc, cactusADM, namd, soplex, sjeng, and libquantum.

In our setup, performance counters are sampled and reset
every 100k cycles. The thresholds for counter selection discussed in
Section~\ref{subsec:counter_selection} are $\alpha = 0.7$ and
$\beta = 0.95$, and were empirically determined.

\begin{scriptsize}
  \begin{table*}[!htb]
  \footnotesize
    \centering
    \caption{Performance bug types injected in gem5 and their
      corresponding locations. Multiple variations of each type were
      implemented for this evaluation.}
    \label{tab:bugs_gem5}
    \resizebox{0.99\textwidth}{!}{
\begin{tabular}{p{0.1\textwidth}|p{0.9\textwidth}}
  \hline
  \textbf{Bug location} & \textbf{Bug type description} \\ \hline
  \multicolumn{1}{l|}{\multirow{2}{*}{Fetch}} & Fetching instructions from the instruction cache takes \textit{T} cycles longer than expected. \\ \cline{2-2} 
  \multicolumn{1}{l|}{} &  Every \textit{T} cycles, the maximum number of instructions that the processor is able to fetch is reduced by \textit{N} entries during one cycle. \\ \hline
  \multicolumn{1}{l|}{Decode} & Instructions that require no source operands are delayed by \textit{T} cycles on the decode stage. \\ \hline
  \multicolumn{1}{l|}{\multirow{5}{*}{Issue}} & If an instruction with opcode \textit{X} reaches the front of the instruction queue, meaning that it has become the oldest instruction there, then the issue process is stalled until the instruction can be issued (all the dependencies have been met, and computational resources are available). Once this occurs, only that instruction leaves the queue during that cycle. Normal behavior is resumed afterward. \\ \cline{2-2} 
  \multicolumn{1}{l|}{} & Every instruction whose opcode is \textit{X} can be retired from the instruction queue only when it becomes the oldest instruction there. A similar bug was found in the Intel Xeon Processors errata~\cite{intel_xeon_errata}. Its description can be found under ``POPCNT instruction may take longer to execute than expected''. \\ \cline{2-2} 
  \multicolumn{1}{l|}{} & If the operands of instructions with opcode \textit{X} depend on the result of an instruction with opcode \textit{Y}, the issuing of the former is stalled by \textit{T} cycles after its operands are ready. \\ \cline{2-2} 
  \multicolumn{1}{l|}{} & If less than \textit{N} slots are available in the instruction queue, delay the next instruction by \textit{T} cycles. \\ \cline{2-2} 
  \multicolumn{1}{l|}{} & The pointer signaling the front of the instruction queue is randomly shifted by \textit{N} positions. This event occurs with a frequency of \textit{T} times per 1000 cycles. \\ \hline
  \multicolumn{1}{l|}{Rename} & All instructions whose opcode is \textit{X} are marked as serializing instructions. This causes all subsequent instructions to be stalled until that instruction has been issued. \\ \hline
  \multicolumn{1}{l|}{\multirow{3}{*}{Execute}} & The latency of functional units handling integer operation \textit{X} is increased by \textit{T} cycles. \\ \cline{2-2} 
  \multicolumn{1}{l|}{} & The latency of functional units handling floating-point operation \textit{X} is increased by  \textit{T} cycles. \\ \cline{2-2} 
  \multicolumn{1}{l|}{} & The latency of functional units handling  ``Single-Instruction Multiple-Data'' operations  \textit{X} is increased by \textit{T} cycles. \\ \hline
  \multicolumn{1}{l|}{Branch} & Branch prediction index table malfunction, effectively reducing the size of the tables by \textit{N} entries. \\ \hline
  \multicolumn{1}{l|}{\multirow{3}{*}{Registers}} & If an instruction with opcode \textit{X} uses physical register \textit{R}, then this instruction is delayed by \textit{T} cycles.  A bug similar to this can be found on Intel 386 DX errata~\cite{intel_386_errata} labeled as ``POPA/POPAD instruction malfunction''. \\ \cline{2-2} 
  \multicolumn{1}{l|}{} & After every \textit{N} times a register has been written, delay the following write by \textit{T} cycles. The inspiration for this bug is the one labeled as ``GPMC may stall after 256 write accesses in NAND\_DATA, NAND\_COMMAND, or NAND\_ADDRESS'' found on the TI AM3517 and TI AM3505 ARM processors errata~\cite{TI_am3517_errate}. \\ \cline{2-2} 
  \multicolumn{1}{l|}{} & The number of physical registers is reduced by \textit{N}. \\ \hline
  \multicolumn{1}{l|}{\multirow{2}{*}{Load/Store Queue}} & For every \textit{N} requests, the load-queue incorrectly rejects entries stating that it is full. \\ \cline{2-2} 
  \multicolumn{1}{l|}{} & For every \textit{N} requests, the store-queue incorrectly rejects entries stating that it is full. \\ \hline
  \multicolumn{1}{l|}{\multirow{2}{*}{Memory}} & After every \textit{N} stores to the same cache line, delay the following write by \textit{T} cycles. \\ \cline{2-2} 
  \multicolumn{1}{l|}{} & The latency of L2 cache is \textit{T} cycles higher than expected. This issue is similar to the documented for NXP MPC7448 RISC processor in its errata~\cite{nxp_7448_errata} labeled as ``L2 latency perfomance issue''. \\ \hline
  \multicolumn{1}{l|}{Re-Order Buffer} & If less than \textit{N} slots are available in the re-order buffer, delay the next instruction by \textit{T} cycles. \\ \hline
  \multicolumn{1}{l|}{Commit} & Every \textit{T} cycles, the maximum number of instructions that the processor is able to commit is reduced by \textit{N} entries during one cycle. \\ \hline

\end{tabular}
}
\end{table*}
\end{scriptsize}

\subsection{Simulated Architectures} \label{subsec:sim_archs}

The experiments shown in Section~\ref{sec:evaluation} are based on
simulations performed in gem5~\cite{gem5} using the out-of-order core
model (O3CPU) and x86 ISA in system emulation mode.
%Although all the
%evaluations are based on simulations, the methodologies can also be
%applied in post-silicon debugging with minor modifications, as discussed in Section~\ref{sec:scope}.

We configured gem5 to emulate a diverse set of microarchitectures that
are used to train and test our scheme. The modified core related settings
are: clock period, pipeline width, branch
predictor, and the sizes of re-order buffer, load/store queue, and
instruction queue.  Cache related configurations (size, associativity,
latency, and number of levels), as well as functional unit
characteristics (count, latency, and port organization) are also
modified to map the simulator behavior to the emulated core.  Other
microarchitectural differences besides these are not considered.

In total, 23 different microarchitecture configurations were
implemented, 15 of them are based on multiple Stock-Keeping Unit (SKU)
variants of the Intel Core microarchitectures: Sandy Bridge, Ivy
Bridge, Haswell, Broadwell, Ice Lake and Skylake. The remaining
were artificially created, but use realistic settings.

For \textbf{training} purposes, we use the data from simulations of the eight artificial
microarchitectures, and the SKU variants of Sandy Bridge and Skylake.
To \textbf{test} the techniques we use the remaining four microarchitectures. As such, the test
microarchitectures are not used to train the ML models.

\subsection{Implemented Bugs} \label{subsec:impl_bugs}

Given its wide acceptance on the computer architecture community, the
gem5~\cite{gem5} simulator is treated as a performance bug-free
simulator. Therefore, in order to evaluate the methodologies,
performance bugs are artificially injected into it. 
Although we use gem5 as a ``bug-free'' baseline, just as any big
software (or hardware) project, it is likely to have bugs,
we believe this should not deter us from implementing debugging 
mechanisms. 

To obtain a list
of bugs that cover most microarchitectural units and are considered
realistic, multiple errata of commercial microprocessors were
reviewed~\cite{intel_xeon_errata,nxp_7448_errata,TI_am3517_errate,intel_386_errata}
and industry experts were consulted. Ultimately, a list of 22
different bug types were implemented in gem5 and are used to evaluate
our scheme. To implement these bugs, we manually edit gem5 code
such that the simulator would behave as the bug describes,
rather than follow its normal behavior.

Although relatively few in the scope of all possible bugs,
these bugs represent a reasonable start for early work in the area.
We examine our localization technique by localizing the
bug to one of 11 possible units. The assignment of bug locations
to each unit, and the description of each performance bug type are
found in Table~\ref{tab:bugs_gem5}. Since the evaluation of the
techniques is based on simulations, no further breakdown is
possible in most units, given the abstraction of unit implementation
in gem5.

\begin{figure}[htb!]
  \centering
  \includegraphics[width = 0.32\textwidth]{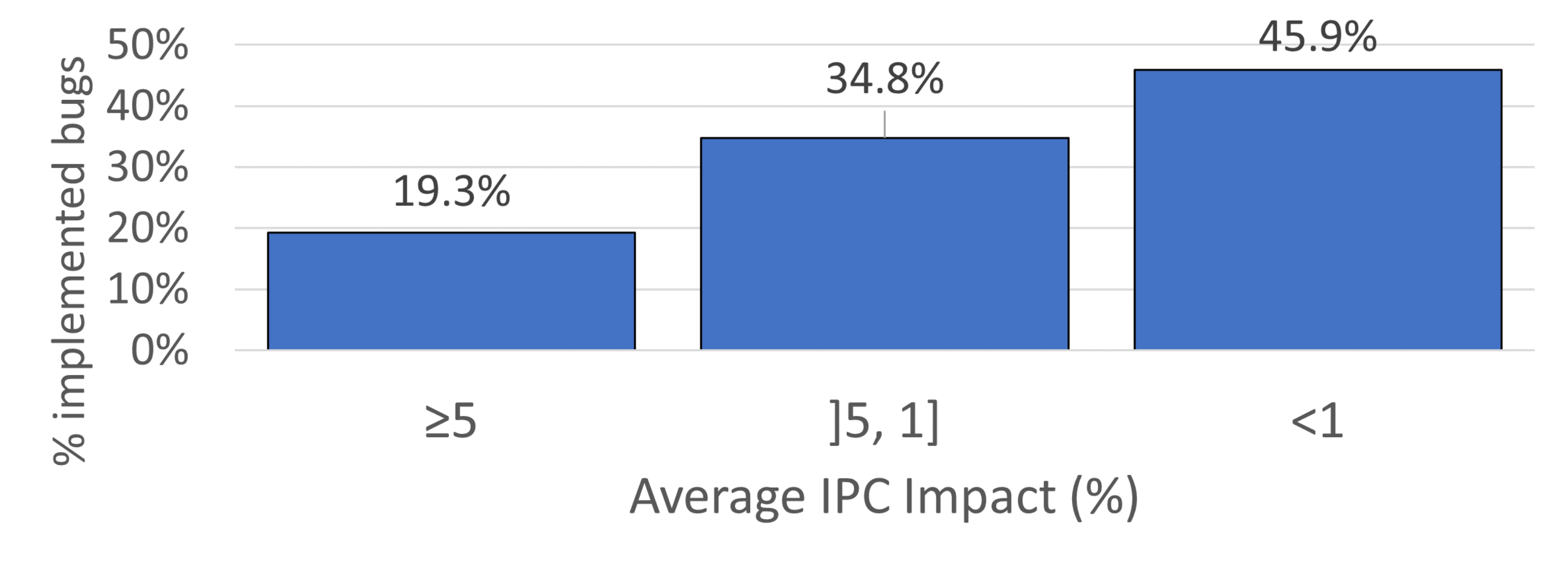}
  \caption{Average IPC impact distribution for evaluated bugs.}
  \label{fig:bug_hist}
\end{figure}

For each of these bug types, multiple variations were implemented by
modifying the values of \textit{X}, \textit{Y}, \textit{N}, \textit{R}, and \textit{T}.  The
average IPC impact of these bugs is measured across the used SPEC
CPU2006 applications, and its distribution is shown in
Figure~\ref{fig:bug_hist}.  Bugs with large IPC impact ($>5\%$) are
usually easier to debug, therefore, we include a smaller fraction of
these bugs.  Bugs that produce a performance degradation between
1\%$-$5\%, can be considered to produce a moderate impact, while
degradation $<$1\% is considered small.  Only bugs with IPC degradation $>0.1\%$ were considered for our
evaluations.

For some bug types, none of their variations are used to train the ML models,
so the methodology can be evaluated on bugs that it has
never encountered before, we refer to these as the ``unseen bug
types''.  For the other bug types, some  variations are used to
train the ML models, while others are left exclusively for testing. This partitioning
evaluates how the model performs on bugs that are similar, but not equal, to
those that it has encountered before
(with the variations left exclusively for testing), as well as in cases where
the same bug seen in legacy designs is encountered (with the variations used for training). We refer to these
as ``unseen variations of seen bug types'' and ''seen bug
variations'', respectively. 
For example, consider a particular bug type X, for which we have three variations: X.A, X.B and X.C. If we use data from legacy architectures
with X.A and X.B to train our models, a sample from a test architecture with X.C would
be an ``unseen bug variation of a seen bug type'' while samples from test architectures
with X.A and X.B would be from a ``seen bug variation''. On the other hand, if none of the training samples
have bug X, the samples from test architectures with any variation of this bug would be considered to be from an
``unseen bug type''.

We collected data for three different variations of each of the 22 implemented bug types. For six of those,
neither variation was used during training (``unseen bug types''). For the remaining 16 bug types,
two variations are used for training (``seen bug variations''), while the other one is left exclusively 
for testing (``unseen bug variation of a seen bug type''). In summary, 27.3\% of the samples correspond to
``unseen bug types'', 24.2\% correspond to ``unseen bug variation of a seen bug type'' and 48.5\% correspond to ``seen bug types''.

The data organization is as follows:

\begin{compactitem}
  \itemsep0em 
\item Training data:
  \begin{compactenum}
    \itemsep0em 
  \item Data with positive labels: For each model trained to identify
    performance bugs in unit $u_j$, the samples with positive labels
    are those from bugs in unit $u_j$ which are placed in
    the ``seen bug variations'' category. These are
    exclusively from the ``train'' microarchitectures.
  \item Data with negative labels: For each model corresponding to
    unit $u_j$, the samples with negative labels are those from
    bugs that do not occur in unit $u_j$ 
    (or bug-free cases, if available) and 
    which are
    placed in the ``seen bug variations'' category. These include data
    exclusively from the ``train'' microarchitectures.
  \end{compactenum}
\item Testing data:
  \begin{compactenum}
    \itemsep0em 
  \item Designs with bugs: Samples from ``seen'' and the two
    ``unseen'' categories of bugs coming exclusively from ``test''
    architectures are evaluated.  In any given sample, only one bug is
    inserted.
  \item Bug-free designs: As mentioned earlier, it is assumed that a
    performance bug has already been detected on the design under
    test. However, an analysis of bug-free
    architectures using the methodologies is shown in
    Section~\ref{subsec:no_bug_handling} to evaluate
    false-positives at detection.
  \end{compactenum}
\end{compactitem}

Note that all the samples used for training are
taken from ``seen bug variations'' in ``train'' architectures. On the
other hand, every sample used for testing comes from ``test''
architectures but can either be a ``seen bug variation'', an
``unseen variation of a seen bug type''  or a
completely ``unseen bug type''.

For our evaluation, at most one bug is injected in every design. Although this is a gap to be addressed in future work, the assumption of a single dominant bug is a first step towards solving the problem. We expect that our methodology would work with multiple bugs because, given that the classifiers for each unit are independent (due to OvA), our methodology should produce high confidence of bugs in all (or most) of the units where a bug is detected. 
\subsection{Bug Localization in Memory Systems} \label{subsec:setup_memory}

Although the focus of this work is the localization of performance
bugs in microprocessor cores, we evaluate the methodologies on
the cache memory system, in order to determine how the methodologies
perform in different setups. Both methodologies are
applied in the exact same manner, but there are minor differences in
the setup used for this evaluation.

Instead of gem5, the ChampSim~\cite{champsim} simulator is used, as it
provides a detailed memory system simulation with a much shorter
simulation time.  Further, by using ChampSim, we highlight the
robustness of our proposed approach.  A total of 96 SimPoints
extracted from 20 applications from the SPEC CPU2017~\cite{spec2017}
benchmark suite are used.  These traces were obtained from the Third
Data Prefetching Championship~\cite{dpc3}. Each of these SimPoints is
2B instructions long, but simulations are stopped after 1B
instructions have been executed. The performance counters are sampled
every 500k cycles due to the long traces being used.

\begin{scriptsize}
\begin{table}[!htb]
  \footnotesize
\centering
\caption{Performance bugs injected to ChampSim and their corresponding locations.}
\label{tab:bugs_champsim}
\resizebox{0.48\textwidth}{!}{%
\begin{tabular}{p{0.12\textwidth}|p{0.35\textwidth}}
  \hline
  \textbf{Bug location} & \textbf{Bug description} \\ \hline
  \multirow{2}{*}{Replacement Policy} & During a cache eviction, the policy evicts the most recently used block, instead of the least recently used. \\ \cline{2-2} 
                        & When a cache block is accessed, the age counter for the replacement policy is not updated. \\ \hline
  \multirow{2}{*}{Prefetcher} & On lookahead prefetching, the path with the least confidence is selected. \\ \cline{2-2} 
                        & Signature Path Prefetcher (SPP)~\cite{kim2016spp} signatures are reset, making the prefetcher use the wrong address. \\ \hline
  \multirow{3}{*}{Other  Operations} & After \textit{N} load misses on L1 data cache, the following L1 data read operation takes  \textit{T} additional clock cycles. \\ \cline{2-2} 
                        & After \textit{N} load misses on L2 cache, the following L2 read operation takes  \textit{T} additional clock cycles. \\ \cline{2-2} 
                        & If there are more than  \textit{Y} misses in less than  \textit{X} cycles, every read operation is delayed by  \textit{T} clock cycles. \\  \hline

\end{tabular}%
}
\end{table}
\end{scriptsize}

The emulated architectures are Intel Nehalem, Sandy Bridge, Ivy
Bridge, Haswell and Skylake, as well as AMD K10 and Ryzen7, and four
artificially generated configurations.  Ryzen7, Haswell and Skylake
are used as testing architectures, while the rest are used for
training the models.

The description of the performance bugs injected to ChampSim, along
with their corresponding locations on the design can be found in
Table~\ref{tab:bugs_champsim}.  Due to the limited number of bugs
available all the bugs are included in the ``seen'' bugs set.

\begin{figure*}[!ht]
    \centering
     \begin{subfigure}[b]{0.285\textwidth}
     \includegraphics[width=0.98\textwidth]{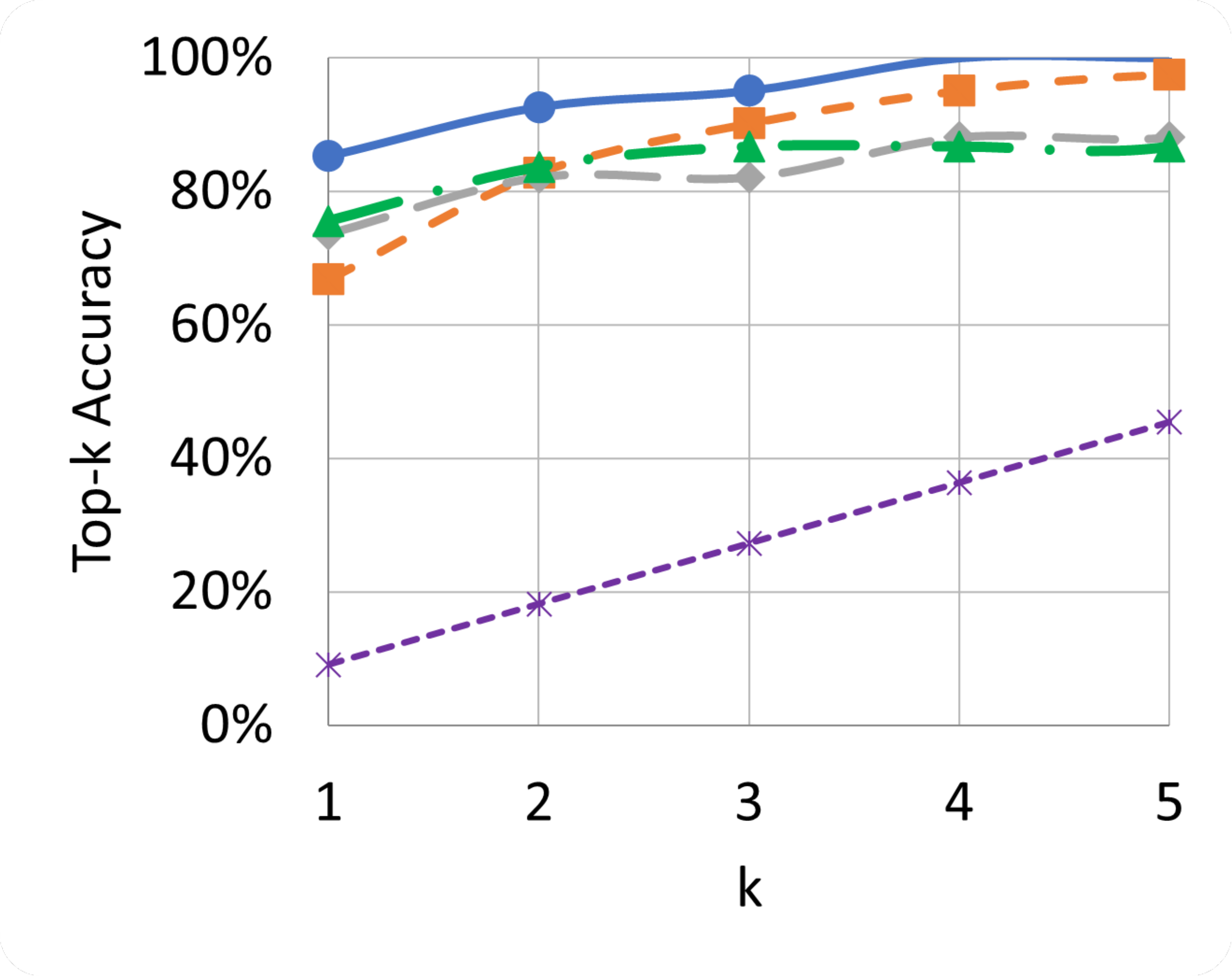}
       \caption{Unseen variations of seen bug types.}
        \label{fig:greater_1_var}
    \end{subfigure}
    ~ 
    \begin{subfigure}[b]{0.285\textwidth}
     \includegraphics[width=0.98\textwidth]{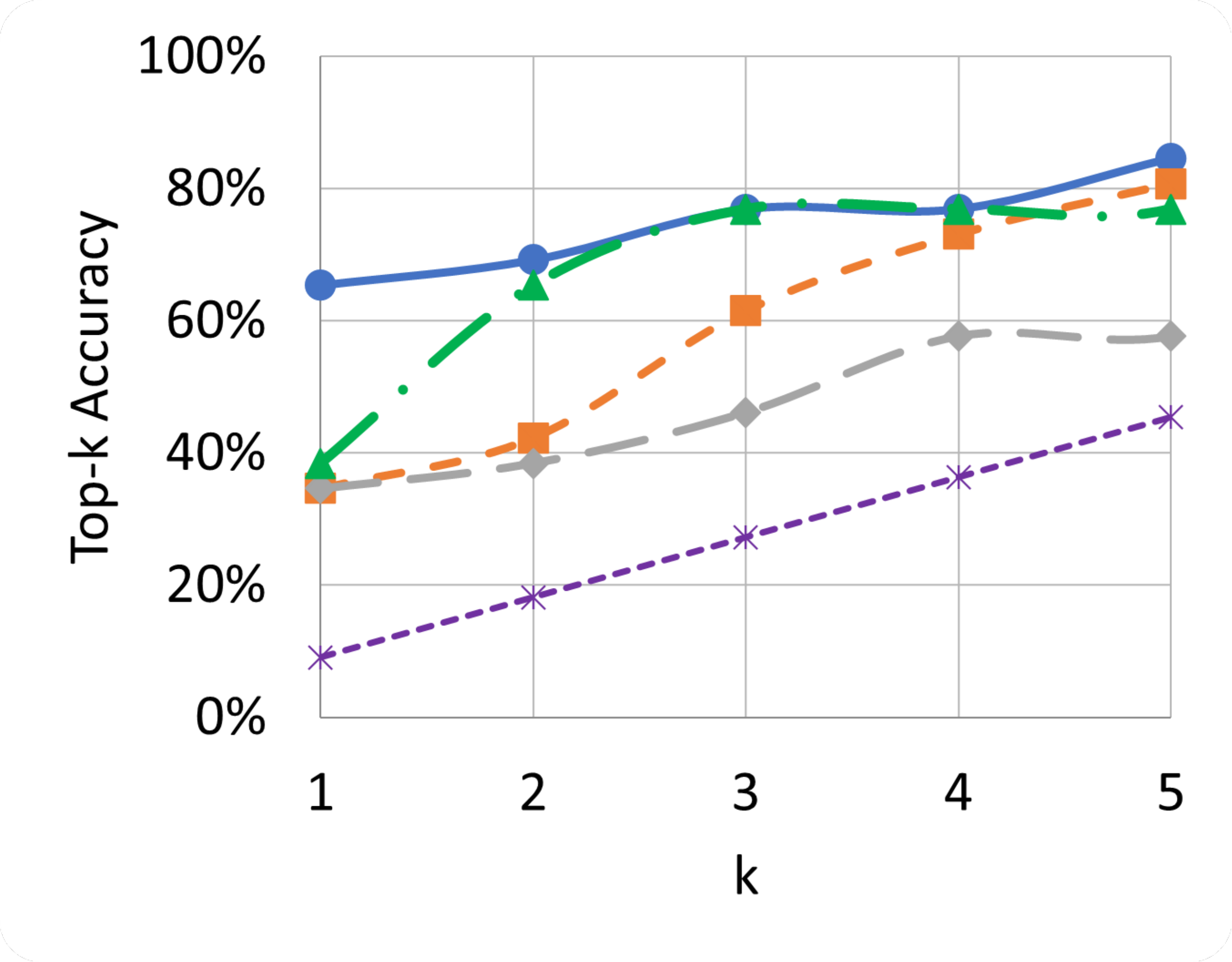}
       \caption{Unseen bug types.}
        \label{fig:greater_1_type}
    \end{subfigure}
    ~
    \begin{subfigure}[b]{0.38\textwidth}
     \includegraphics[width=0.98\textwidth]{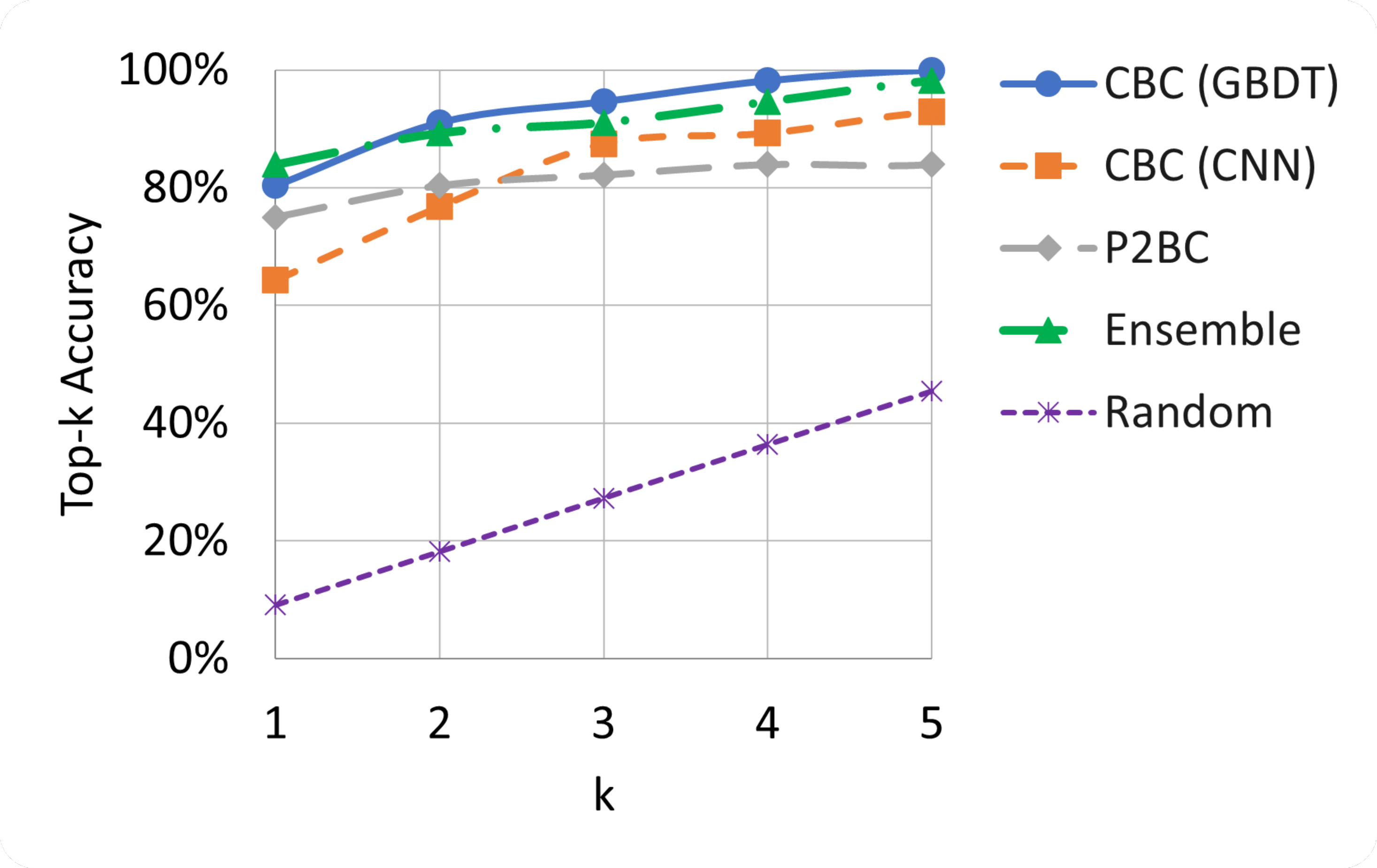}
        \caption{Seen bug variations.}
        \label{fig:greater_1_trained}
    \end{subfigure}
    \vskip 1ex
    \caption{Top-\emph{k} accuracy for bugs with average IPC impact $>$1\%.}
    \vskip 1ex
    \label{fig:per_ipc_1}
\end{figure*}

\section{Evaluation} \label{sec:evaluation}

This section presents the details of the experiments conducted to
evaluate our methodologies.  Sections~\ref{subsec:per_impact_accuracy}
through~\ref{subsec:no_bug_handling} show the results of the
methodologies when applied to the microprocessor core setup.
Section~\ref{subsec:memory_accuracy} presents the results of the
examination on memory systems.  Since we are not aware of prior work
in automatic performance bug localization, no comparison to prior work is
conducted.

Since the localization problem is formulated as a multi-class
classification task, both methodologies produce a sorted list from
highest to lowest probability of the performance bug being located in
each unit. The top-\emph{k} accuracy metric is used to measure results
quality. Here, a result is considered correct if the actual bug
location is found among the first \emph{k} choices suggested by our
techniques. The reason why top-\emph{k} accuracy is more relevant in
this work than other metrics (\emph{e.g.} a confusion matrix) is
because, rather than measuring when a sample is incorrectly classified
into a different class, it is more important to determine how high in
the predicted ranking the actual bug location is, as this represents
how many units the designer or validation engineer would have to search to actually find the
bug. Ideally, we would like the methodology to pinpoint to a single
unit (top-1 accuracy), but due to the challenging nature of the problem, we find 
top-k accuracy a good compromise.

Note that the accuracy of a random guess in a multi-class task is not
50\%, as in the more frequently studied binary classification, instead,
it is actually $1/|U|$ where $U$ is the set of all classes
(units). Therefore, the top-\emph{k} accuracy of a random guess is
given by $k/|U|$. The reported accuracy is obtained as the average
across the four architectures used as test cases.

It is important to note that bugs with small
average IPC impact are hard to catch, and do not represent a priority
for the designers as the gain by fixing them is not high.
Therefore, our main results, presented in Section~\ref{subsec:per_impact_accuracy}, focus in the 
analysis of the results obtained
on bugs with average IPC impact  $>$1\%. Section~\ref{subsec:overall_accuracy} shows the performance
of our methodology when smaller bugs (average IPC impact $>$0.1\%.) are considered.

All the methodologies were implemented in Python. For neural networks
Keras~\cite{chollet2015keras} is used, and for the gradient boosted decision
trees, XGBoost~\cite{chen2016xgboost} is used.

\subsection{Accuracy on High Impact Bugs} \label{subsec:per_impact_accuracy}

In this section, we show the top-\emph{k} accuracy across the implemented bugs
with an average IPC impact $>$1\% in our test architectures. Figure~\ref{fig:per_ipc_1} shows
the top-\emph{k} accuracy obtained by our proposed methodologies as the value of \emph{k}
increases. These results are separated using the bug partitioning scheme explained in Section~\ref{subsec:impl_bugs}.
Figure~\ref{fig:greater_1_var} shows the behavior across the 
``unseen variations of seen bug types'',
Figure~\ref{fig:greater_1_type} shows the results on the 
``unseen bug types'' and Figure~\ref{fig:greater_1_trained} shows the same, but
for the ``seen bug variations'' exclusively.  The figures show values of
\emph{k}$<$5, as longer lists of possible locations do not
provide significant help to designers to conduct their debug.

In the figures, ``CBC (GBDT)'' refers to the case of CBC with
per-time-step classification and gradient boosted tree models 
(100 trees per model, other parameters left at default).
The ``CBC (CNN)'' performs a per-trace classification using
convolutional neural networks, as described in
Section~\ref{subsec:one_stage_methodology}
(2 1D-CONV layers with 100 filters each, followed by
3 FC layers with 300, 100, 50 neurons with ReLU activation functions 
and a final output layer of a single neuron with sigmoid activation).
The ``P2BC'' results
correspond to an implementation with CNNs for the classifiers on the
second stage (the first stage uses GBDT with 250 trees, while the second
stage uses the same architecture as ``CBC (CNN)''). 
The results labeled as ``Random'' represent what would
be obtained with a random guess. The baseline for comparison should be the 
state of the art, which is manual debug. However, quantifying manual debug is difficult, as it varies significantly from person to person. We consider a random result a practically quantifiable baseline for comparison. Even if a manual debug is more accurate than random, given that our methodologies are automated, we are able to speed up the process significantly vs manual debug.

In this case, CBC (GBDT) is the best performing technique, even better
than the ensemble. The accuracy obtained by the ensemble is
impacted due to P2BC having inferior accuracy for higher impact bugs
when compared to CBC (GBDT). 

Considering all the bugs (weighted average across the three bug partitions), our CBC (GBDT) methodology is able to achieve 76.8\% top-1 accuracy and over 90\% top-3 accuracy. As shown in  
Figure~\ref{fig:per_ipc_1}, results for ``unseen variations of seen bug types''
and ``seen bug variations'' are above that, both achieving a top-3 accuracy of about 95\%. As expected, the
``unseen bug types'' are the hardest to predict accurately, however, even
for these difficult cases of bugs that do not look similar to the ones
used for training, our CBC (GBDT) methodology achieves 76.9\%
top-3 accuracy.

The performance impact produced by a 
performance bug is not necessarily observed on the unit producing the bug. 
Our methodologies make no assumptions regarding this, in fact, the bug which incorrectly marks instructions as “serialized” is an example of how our methodologies are able to handle this. Our methods identify this type of bugs as occurring in the “Rename” stage, however serializing instructions will have an impact on the performance starting at the “issue” stage.

In our evaluation we found the units ``Branch'' and ``Memory'' 
to be the only ones with a localization rate much lower than the average. However, this is also
correlated with the fact that the implemented bugs for those two units had a much lower IPC
impact than the bugs on the other units.

\begin{figure*}[h!]
  \centering
  \begin{subfigure}[b]{0.285\textwidth}
    \includegraphics[width=0.98\textwidth]{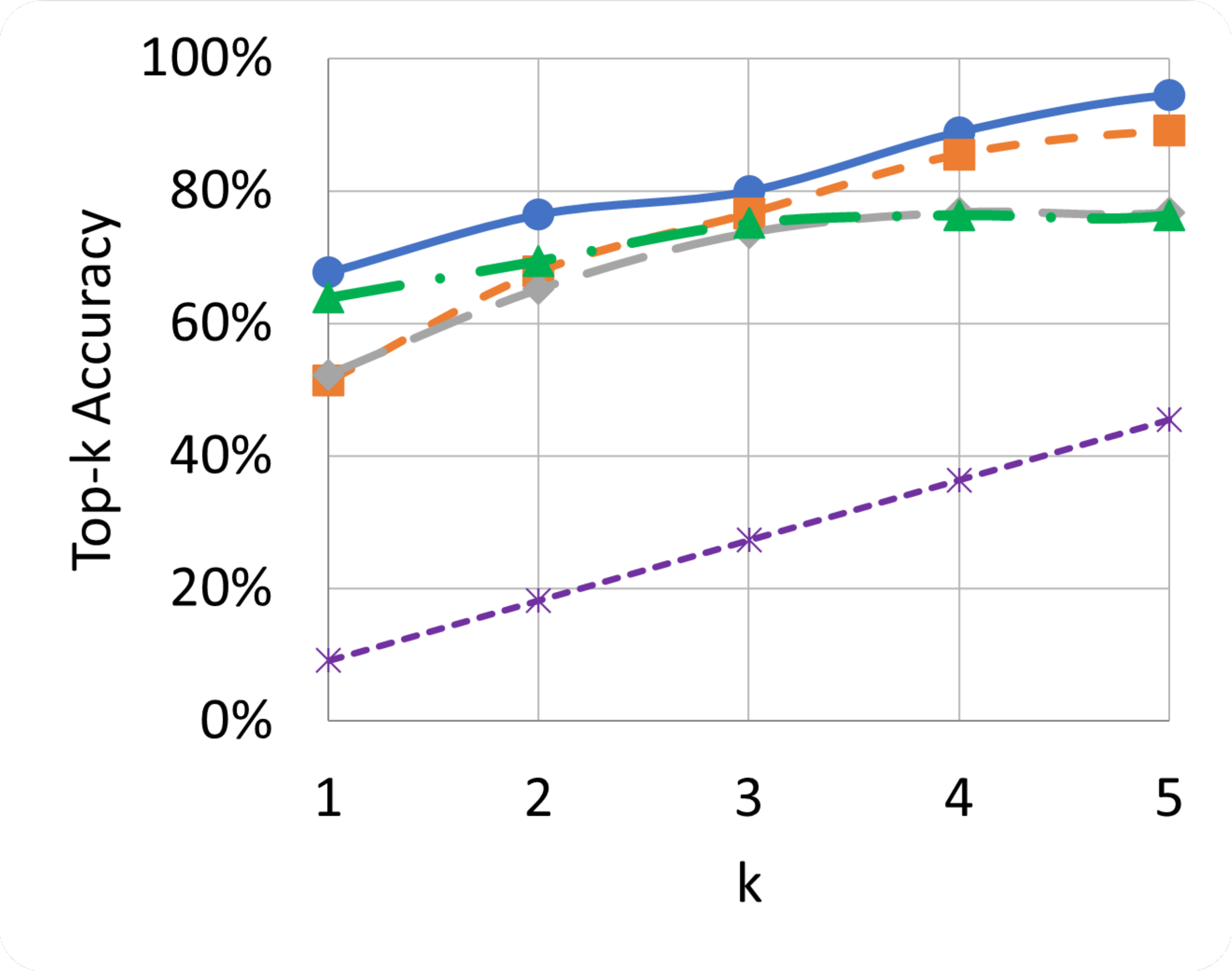}
    \caption{Unseen variation of seen bug types.}
    \label{fig:unseen_var_acc}
  \end{subfigure}
  ~
  \begin{subfigure}[b]{0.285\textwidth}
    \includegraphics[width=0.98\textwidth]{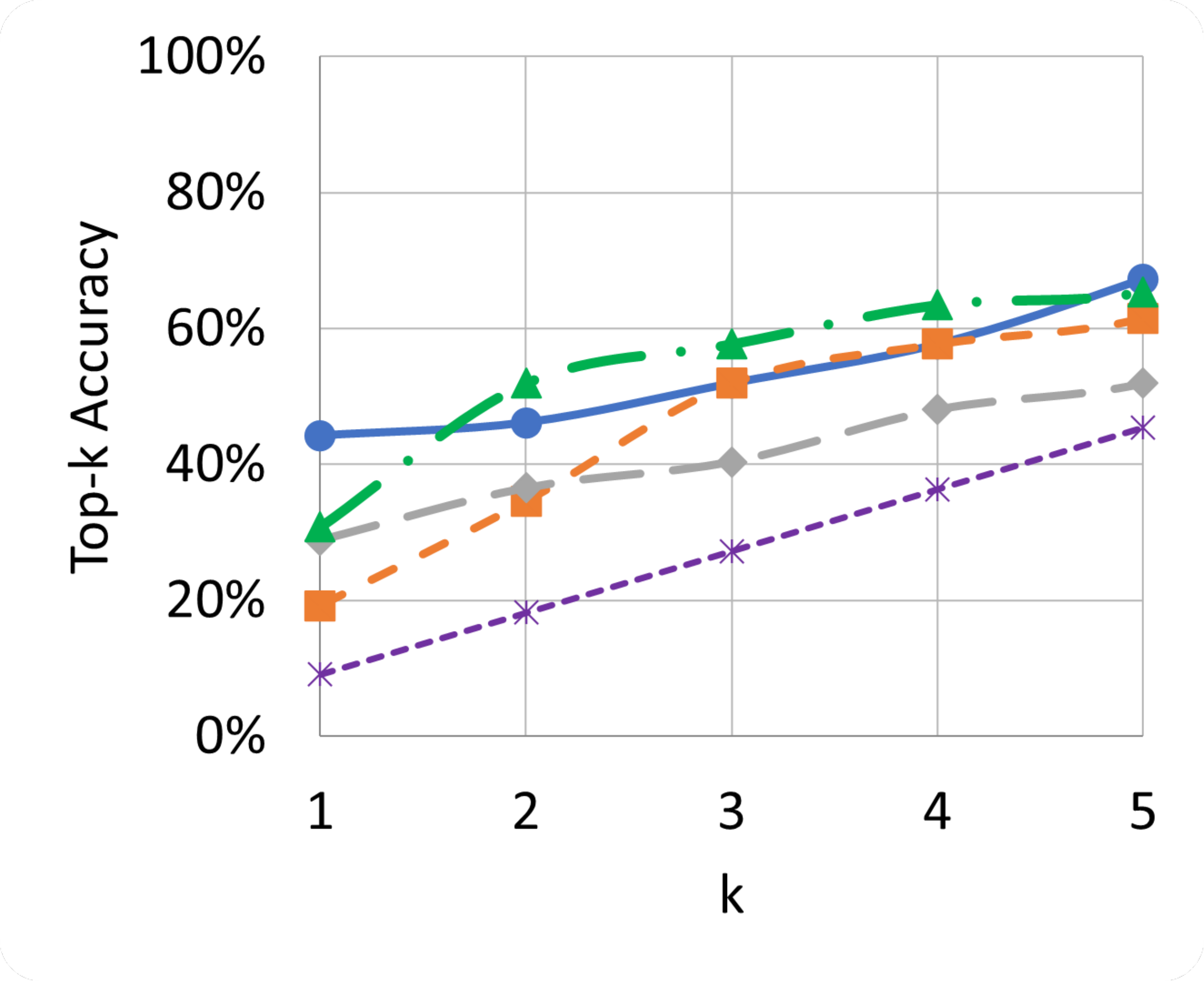}
    \caption{Unseen bug types.}
    \label{fig:unseen_type_acc}
  \end{subfigure}
  ~
  \begin{subfigure}[b]{0.38\textwidth}
    \includegraphics[width=0.98\textwidth]{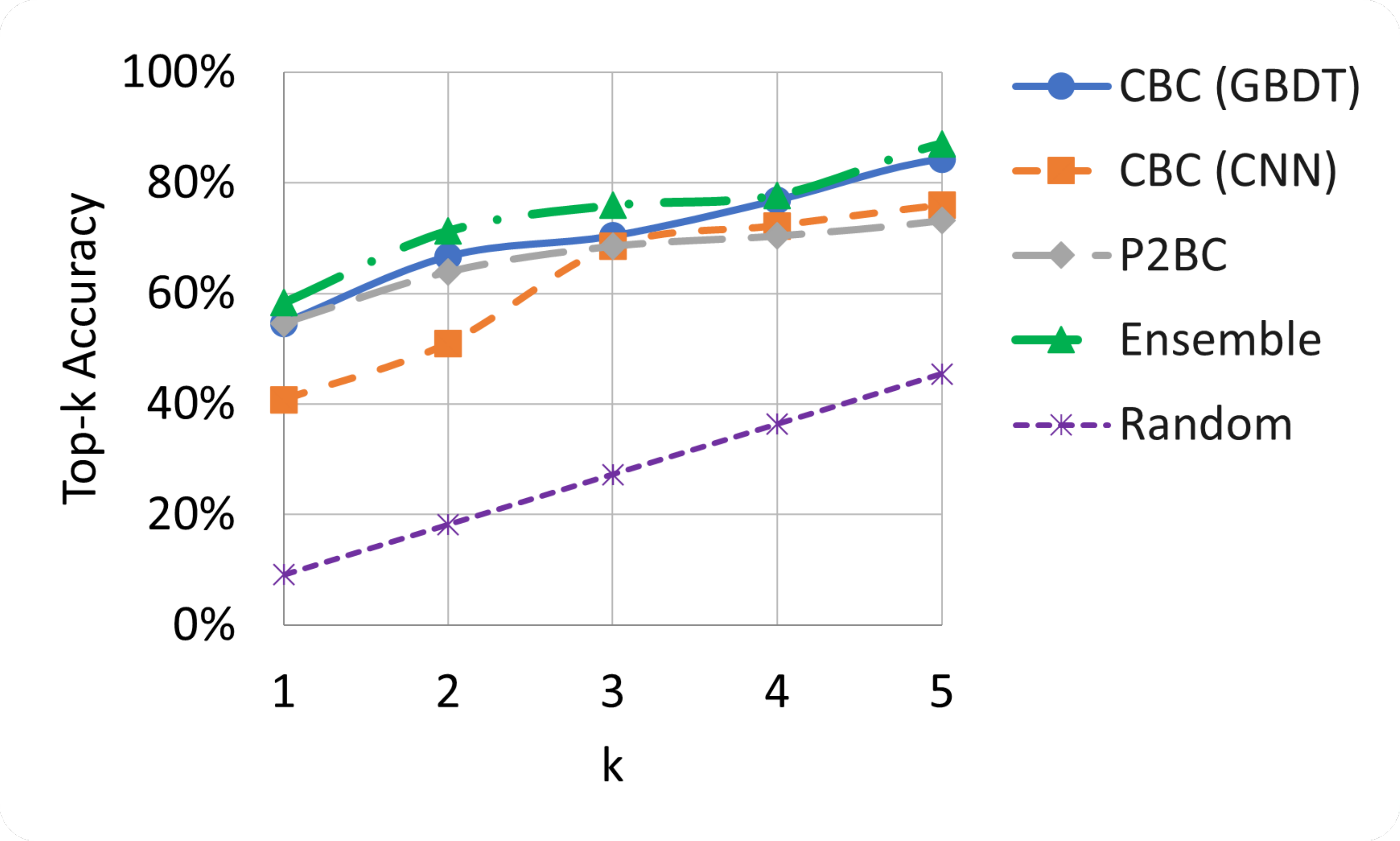}
    \caption{Seen bug variations.}
    \label{fig:seen_acc}
  \end{subfigure}
  \vskip 1ex
  \caption{Top-\emph{k} accuracy for bugs with average IPC impact $>$0.1\%.}
  \vskip 1ex
  \label{fig:top_k_gt01}
\end{figure*}

\subsection{Accuracy on Low Impact Bugs} \label{subsec:overall_accuracy}

In this section, we analyze the accuracy of our methodologies
for bugs with smaller average IPC impact ($>0.1\%$).
The results for this case are shown in Figure~\ref{fig:top_k_gt01}.

Overall, across all the evaluated bugs, the Ensemble method provides
the highest top-3 accuracy with a 72.5\%. The advantages of the Ensemble
are prominent on the cases of ``unseen bug types'' (Figure~\ref{fig:unseen_type_acc}) and ``seen bug variations'' 
(Figure~\ref{fig:seen_acc}). In this case, just as observed in
Section~\ref{subsec:per_impact_accuracy}, the ``unseen bug types''
are the hardest to localize accurately. Although CBC (GBDT) performs
slightly better than the Ensemble for ``unseen
variations of seen bug types'' (Figure~\ref{fig:unseen_var_acc}), when
all the cases are considered, the Ensemble methodology outperforms it in
this average IPC impact.

As it can be observed by contrasting Figures~\ref{fig:per_ipc_1} and~\ref{fig:top_k_gt01}, the results degrade when bugs with
smaller impact are considered. However, we believe that the our 
results 
are satisfactory given that a 0.1\% IPC impact can be negligible in most
situations.

When each technique is used individually, CBC~(GBDT) is the
best performing, being able to identify the correct location of the
bug in the top-3 ranked options in around 80\% of the cases when a 
similar bug (but not the exact same) was used for training the models.
Although the accuracy achieved by P2BC is not as high as the obtained
by CBC, it provides satisfactory results while using $100\times$ less
storage.
\subsection{Number of Workloads} \label{subsec:nb_probes}

\begin{figure}[htb!]
  \centering
  \includegraphics[width = 0.37\textwidth]{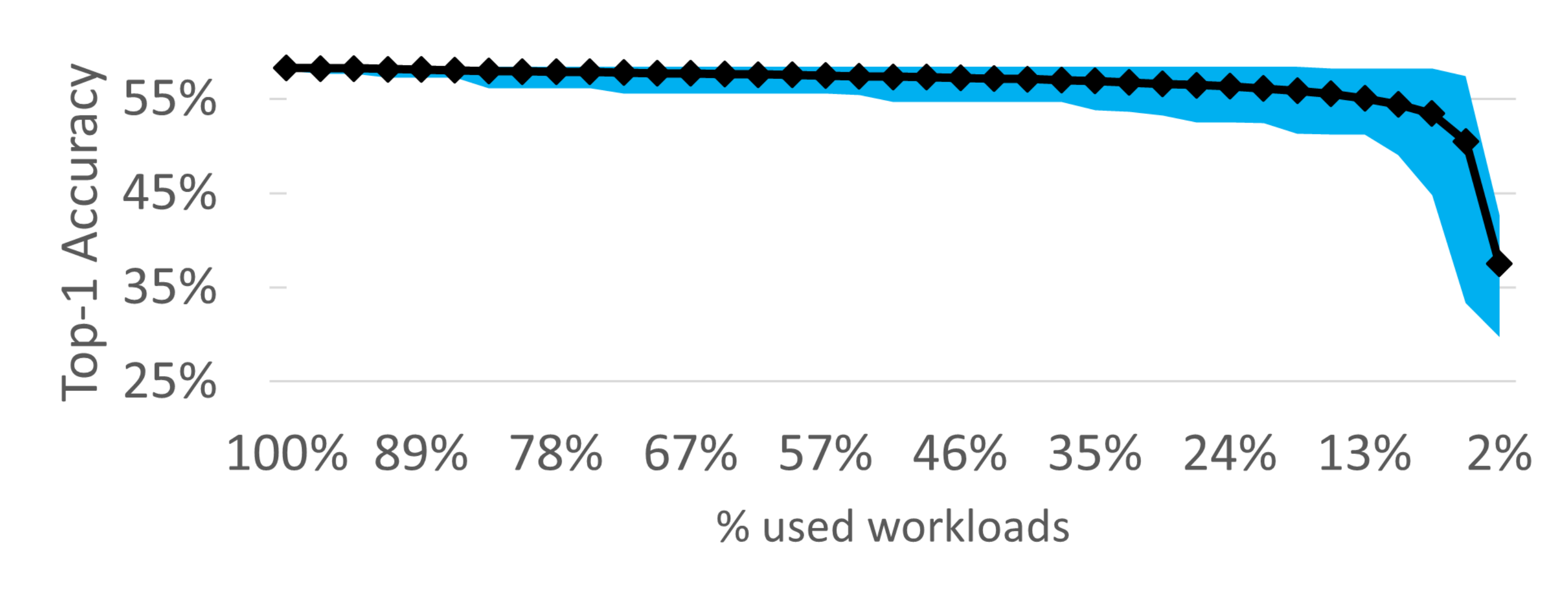}
   \caption{Impact of the number of workloads in the top-1 accuracy across all bugs.}
  \label{fig:probe_coarse}
\end{figure}

In this section, the impact of the number of workloads on the 
method's accuracy across bugs with average IPC impact greater than 0.1\%
is evaluated.  This evaluation was conducted for the ``CBC
(GBDT)'' implementation, as it was found to be the best performing
stand-alone method.  The experiment starts by using the 190 workloads available
for this work and measuring the top-1 accuracy obtained. The number of 
workloads is iteratively reduced by randomly choosing five to be
discarded in each iteration.  The iterations continue until only five
workloads remain.  This experiment is repeated 100 times to reduce the
impact of the random choices of workload deletion and obtain a
reliable trend.  Figure~\ref{fig:probe_coarse} shows the obtained
results.

In the figure, the black line represents the average top-1 accuracy
across the 100 repetitions, while the maximum and minimum accuracy of
any individual iteration are represented by the shaded area.  The
figure shows that, as the number of workloads used to test the
methodology decreases, so does the average quality of results, making
a point that as more workloads are available, higher accuracy will be
achieved. However, we note the degradation in the overall accuracy is
slow and proves that the methodology can achieve satisfactory results
even with a reduced number of workloads.
\subsection{Behavior When Design is Bug-Free} \label{subsec:no_bug_handling}

As mentioned in previous sections, the evaluation of this work assumes
that the presence of a performance bug has been detected.  However,
detection methodologies may still produce false-positives \emph{i.e.}
cases where a bug-free design is classified as buggy. In this section,
an analysis of the behavior of both methodologies when such case
occurs is conducted.

\subsubsection{False-Positives in CBC}

Each of the four ``test'' architectures without performance bugs
produce a different ranking of units. However, 
the confidence assigned to the highest ranked unit is
much smaller than the confidence obtained by architectures with bugs.

Although low confidence across all the units may signal a
false positive, it could also indicate a gap of coverage, as there
might be a bug in a unit that is not considered in the evaluated
classes.  To prevent that, a model to detect ``Bug-Free''
architectures is trained in the same way as the models for each possible
bug location unit (\emph{i.e.} a ``Bug-Free'' class is
included to $U$, the list of possible bug locations). Results show that
such model performs very well, it provides the highest ranked
confidence score to the bug-free class in all four bug-free
``test'' architectures. This addition does not degrade
results for buggy samples, as the confidence score for the
``Bug-free'' class is more than $200\times$ smaller than any 1st
choice of all the buggy samples and it is not included as a top-5
option in any sample with a performance bug.

\subsubsection{False-Positives P2BC}

For P2BC, we found that there were multiple units with high
classification confidence. In those cases there was nothing in
particular that allowed to differentiate these bug-free samples from
the average behavior observed in samples with bugs.

Similarly to the evaluated for CBC, a ``Bug-Free'' class addition
was tested. In this case however, the method does not provide a
satisfactory accuracy, making CBC more robust to handle false positive
cases. One possible explanation for this is that P2BC is more sensible to
bugs with small average IPC impact, which makes it more vulnerable
to false positives.

\subsection{Bug localization in memory systems} \label{subsec:memory_accuracy}

For this setup, top-1 accuracy in both methodologies was
of 100\%. Cases of bug-free architecture obtain low confidence scores
in all possible locations, if a model to classify as
``Bug-Free'' using the same methodologies discussed in
Section~\ref{subsec:no_bug_handling} is included, it achieves
100\% detection rate without false negatives. Although this is a small setup for localization of bugs in memory
systems, its results show that the methodologies presented here are
robust for usage in different system components, as well as different
simulation frameworks.

\section{Related Work} \label{sec:related_work}

Although the importance and difficulty of microprocessor performance
debugging has been recognized for over a quarter-century, few works
have attempted to tackle the issue.  Most of these prior works have
focused on detecting whether a performance bug is present in the
design, while only a few have attempted to localize the bugs.

\subsection{Performance Bug Detection}

Bose~\cite{Bose1994} attempted to detect microprocessor performance bugs by building
a performance fault model and simulating application traces.  
Creating this model is a very manual effort, and some of the
decisions taken to simplify the model are not realistic for modern
designs.  Other works used processor timing
models~\cite{surya1994architectural} and developed ad-hoc
microbenchmarks that enforce certain design invariants with repetitive
execution of hand-crafted loops.  Here, a different microbenchmark
must be developed for every test, constraining the coverage
space and, although a failure provides symptoms of a
bug, it does not assert the presence of the bug to any
specific location.

Singhal \emph{et al.}~\cite{Singhal2004} described the procedures and
tools used for performance verification in the Intel Pentium 4
processor on a 90nm technology.  This work shows that detection of
performance bugs relies on the execution of benchmarks and checking
whether the design achieved the expected performance, otherwise, a
manual debug process would be needed.  The recent work by Carvajal
\emph{et al.}~\cite{carvajal2021detection} automated detection
mechanisms with machine learning strategies relying on the knowledge
of performance behavior observed in legacy designs. Although this
strategy shows promise, it is a first step towards automated
performance debugging, as it only shows whether a bug is present in
the entire system, lacking a way to provide details regarding the
location of the detected bug.

\subsection{Performance Bug Localization}

Very few works have attempted microprocessor performance bug localization.  The few
partially related are restricted to test generation, and
none provide an automatic overall methodology.

Utamaphetai \emph{et al.}~\cite{Utamaphethai1999} model the entire
design as buffers and finite state machines in order to use strategies
developed for test generation.  These simplifications make the
methodology unable to localize bugs that are not necessarily related
to state transitions or dependencies.  Adir \emph{et
  al.}~\cite{adir2005generic} perform a microarchitectural-level
verification based on a test plan created by coverage models. These
coverage models however, require ``creativity'' and significant domain
knowledge.  Since the analysis of test program results is still
manual, their evaluation required almost 10 days (without considering
the time to create the coverage models) to validate a single
microarchitectural unit. Scalability concerns can arise when other
units and the interactions between are considered.  Both of these
works focus on test generation techniques while the analysis of test
results is still a conventional manual approach.

\iffalse
A multitude of works have developed
analytical performance models for superscalar out-of-order processors~\cite{noonburg1994theoretical,
karkhanis2004first,eyerman2009mechanistic,clement1993analytical,black1998calibration}
that could be used for performance debugging.
However, in order to achieve high accuracy, a large amount of parameters and combination of 
events that could affect the performance need to be accommodated, which makes analytical models
very complex, hence, although useful, these techniques are not widely used for performance verification. As an
attempt to automate the performance model generation Ould \emph{et al.}~\cite{Ould2007}
developed a tree-based strategy that uses performance counters in order to estimate the
design performance. However, the construction of the model is performed by executing and monitoring
the results of multiple traces on the design for which the model is being built,
which makes this strategy unfit for performance debugging, as the models will reflect
whatever bug present in the design.
\fi

%Others have also evaluated the usage of performance counters as inputs to a model to estimate a target metric has in different areas such as software performance regressions~\cite{shang2015perf} power analysis~\cite{Joseph2001, Contreras2005, Bircher2007,rodrigues2013,tsafack2014exploiting,Yang2016} or functional verification~\cite{BugMD, Poulos2018, Efendioglu2019}.

\subsection{Performance Bugs in Other Domains}

Even though the area of performance debugging in microprocessors has
not received a lot of attention in recent years, several strategies
for performance debugging in other areas have been proposed. The most
widely studied are performance debugging in cloud computing and
software engineering.

In the cloud computing environment, performance issues lie on the
detection of anomalies that affect the quality of the
service~\cite{Ibidunmoye2015}.  Gan \emph{et al.}~\cite{gan2019seer}
developed a deep learning-based online QoS violation prediction
technique for cloud computing by using runtime traces. This work
detects an upcoming violation, and identifies the microservice
responsible for it.  This is similar to the localization task tackled
in this work, but in a different environment where the source
corresponds to a microservice being executed by a system, as opposed
to a microarchitectural unit that is part of the system itself.

Recent works using ML approaches for performance
evaluation in software have been proposed.  Alam \emph{et
  al.}~\cite{alam2019autoperf} detect performance regressions due to
code changes using autoencoders.  In other works, a common strategy
for performance bug localization, is to use a language processing
models based on bug reports from software project
databases~\cite{rao2011retrieval,zhou2012bl, saha2013improving,
  akbar2019scor}.  The models learn to identify the correlations
between specific keywords in bug reports and the code changes
implemented to fix them. Although software performance debugging is
complex, performance debugging in microprocessor is a more complicated
task due to the much larger level of concurrency. 

Other works have attempted automation of functional bug localization~\cite{park2010blog}, 
while interesting, we find that functional bug approaches do not work well for performance 
debugging due to its lack of a known correct baseline. 
\section{Conclusion and Future Work} \label{sec:conclusion}

Here, we presented two automated methodologies for processor
performance bug localization leveraging ML, the first
work of this kind to the best of our knowledge.  Our methodologies
extract information from legacy designs in order to learn the
relationships between performance counter behavior and locations of
performance bugs.  Between the approaches, one is more accurate while
another is $10\times$ faster and has $100\times$ smaller data
footprint.  An ensemble of two of the methodologies was also analyzed, and it was 
found to provide
further accuracy improvement.  Simulation results on the evaluated
bugs show that the top-3 accuracy achieved is as high as 90\% for bugs
with average IPC impact greater than 1\%, and 72\% for bugs with
impact greater than 0.1\%.
%%%%%%% -- PAPER CONTENT ENDS -- %%%%%%%%

%%%%%%%%% -- BIB STYLE AND FILE -- %%%%%%%%
\bibliographystyle{IEEEtranS}
\bibliography{refs}
%%%%%%%%%%%%%%%%%%%%%%%%%%%%%%%%%%%%

\end{document}